\documentclass[3p,11pt]{elsarticle}
\usepackage{amssymb,amsfonts,amsmath,natbib}
\usepackage{lmodern}

\graphicspath{{figures/en/}}

\journal{Journal Title}

\numberwithin{equation}{section}

\begin{document}

\begin{frontmatter}

\title{Exact solutions of the space time-fractional Klein-Gordon equation with cubic nonlinearities using some methods}
\author{Ayten \"Ozkan
\corref{mycorrespondingauthor}}
\author{Erdo\u{g}an Mehmet \"Ozkan}

\begin{abstract}
Recently, finding exact solutions of nonlinear fractional differential equations has attracted great interest.
In this paper, the space time-fractional Klein-Gordon equation with cubic nonlinearities is examined. Firstly, suitable exact soliton solutions are formally extacted by using the solitary wave ansatz method. Some solutions are also illustrated by the computer simulations. Besides, the modified Kudryashov method is used to construct exact solutions of this equation.
\end{abstract}

\begin{keyword}
Space time fractional Klein-Gordon equation \sep ansatz method \sep modified Kudryashov method \sep exact solutions
\MSC[2010] 26A33 \sep 35R11 \sep 83C15
\end{keyword}

\end{frontmatter}
\section{Introduction}
Fractional differential equations are generalization of differential equations . In recent years, non-linear fractional differential equations (FDEs) have achieved importance in various disciplines and have become popular. Recently, the theory and applications of FDEs have been the focus of many studies since they appear frequently in various applications in mathematics, physics, biology, engineering, signal processing, systems identification, control theory, finance, fractional dynamics, and have increasingly fascinated the attention of many scientists. FDEs have been studied and many researchers published books and articles in this field \cite{Miller, Podlubny,Kilbas}. Many methods have been introduced to obtain exact solutions of FDEs. For instance the first integral method \cite{Ray, Taghizadeh, Mirzazadeh, Eslami, Cenesiz}, exp-function method \cite{Guner, Bekir1, Bekir2}, $(G'/G)$ expansion method \cite{Bin, Bekir3, Baleanu}, sub-equation method \cite{Aksoy, Bekir4}, functional variable method \cite{Matinfar, Guner1}, trial equation method \cite{Bulut, Odabasi}. 

A dependable and powerful method called the ansatz method has been put forward to search for traveling wave solutions of nonlinear partial differential equations by Biswas \cite{Biswas1, Karishnan} . Although this method has been used by many authors, the applications of this method are very low in nonlinear FDEs. The installation of exact and analytical traveling wave solutions of nonlinear FDEs is one of the most significant and basic duties in nonlinear science, because they will characterize miscellaneous natural case such as vibrations, solitons and finite speed distribution. The Ansatz method is one of the efficient methods used to obtain exact soliton solutions of FDEs.

The solitary wave study has made important progress recently. In mathematics and physics, a soliton or a solitary wave is a self-reinforcing single wave that moves at a constant velocity, while maintaining its shape. Solitons represent solutions of the class of largely weak nonlinear distributive partial differential equations associated with physical systems. This field of study has recently made a huge progress \cite{Biswas1, Bekir, Bekir5, Bekir6, Biswas, Biswas2, Biswas3, Biswas4, Khalique, Triki}. In the present study, FDEs will be converted into integer-order differential equations by fractional complex transformation, and then various exact solutions will be obtained to determine singular soliton solutions, dark soliton solutions and bright soliton solutions \cite{Guner2, Mirzazadeh1}.

One of the approaches that led to creating exact solutions of fractional differential equations is a modified version of the Kudryashov method \cite{Kudryashov}. The modified Kudryashov method is a powerful solution method for finding exact solutions of nonlinear partial differential equations (PDEs) in mathematical physics and biology. This method was first applied in fractional differential equations by Ege and Misirli \cite{Ege}. Recently, this method has gained considerable attention due to the ability of PDEs to extract new complete solutions both in integer order and in fractional order \cite{Korkmaz, Hosseini, Saha}.

Nonlinear Klein - Gordon equations have important application areas in science and engineering such as solid state physics, nonlinear optics and quantum
field theory \cite{Wazwaz}. This equation is a relativistic field equation for scalar particles and is a relativistic generalization of the well-known Schrödinger equation. Despite other relative wave equations, the Klein-Gordon equation is the most frequently studied equation in quantum field theory, since it is used to describe particle dynamics \cite{biswasA}. They have been studied by many researchers and various methods have been used to solve them. Some of these studies can be listed as follows : Homotopy perturbation method \cite{Alireza}, a semi-analytical method called fractional-reduced differential transformation method with the appropriate initial condition \cite{Tamsir},  modified Kudryashov method \cite{Hosseini1}, fractional complex transformation, $(G '/ G)$ and $(w / g)$ expansion methods \cite{Unsal}, the well-organized ansatz method \cite{Hosseini2}, a direct analytic method \cite{Culha}, the modified expanded Tanh method \cite{Shallal}. In this study, ansatz method and modified Kudryashov method are applied to find out several new explicit exact solutions of the space time-fractional Klein–Gordon equation with cubic nonlinearities.

\section{The modified Riemann-Liouville derivative and methodology of solution}

With recent studies, it is well known that the dynamics of many physical processes are accurately described using FDEs having different kinds of fractional derivatives. The most popular ones are the Caputo derivative, the Riemann-Liouville derivative and Gr\"{u}nwald-Letnikov derivative. A different definition of the fractional derivative is given by Jumarie with a little modification of the Riemann-Liouville derivative. In \cite{Jumarie}, $f:R \rightarrow R$, $\omega \rightarrow f(\omega)$ as a continuous function (not necessarily differentiable), the modified Riemann-Liouville derivative of order $\alpha$  is given as follows
\begin{equation}
D^{\alpha}_{\omega}f(\omega)= \left\{ \begin{array}{ll}
\frac{1}{\Gamma(1-\alpha)} \frac{d}{d\omega} \int^{\omega}_{0}\frac{f(\tau)-f(0)}{(\omega-\tau)^{\alpha}}d\tau & \textrm{ $,0<\alpha<1$, }\\
\\
(f^{(n)}(\omega))^{(\alpha-n)} & \textrm{ $,n\leq \alpha \leq n+1, \ \ n\geq 1 $ }
\end{array} \right.
\end{equation}
where $\Gamma(.)$ is the Gamma function.
In addition, some important properties of the fractional modified Riemann-Liouville derivative (mRLd) are listed as follows \cite{Jumarie1}:
\begin{equation} \label{eq2}
D^{\alpha}_{\omega}\omega^{\gamma}=\frac{\Gamma(1+\gamma)}{\Gamma(1+\gamma-\alpha)} \textrm{ $,\gamma > 1,$ }
\end{equation}
\begin{equation}
D^{\alpha}_{\omega}(c)=0 \textrm{ $ (c  $ constant) $,$ }
\end{equation}
\begin{equation}
D^{\alpha}_{\omega} (a f(\omega)+ b g(\omega))= a D^{\alpha}_{\omega} f(\omega) + b D^{\alpha}_{\omega} g(\omega),
\end{equation}
where $a\neq 0$ and $b \neq 0$ are constants.

\noindent
Now, we will take into account the following nonlinear space-time FDE of the type
\begin{equation} \label{eq4}
H(u,D^{\alpha}_{t}u,D^{\alpha}_{x}u,D^{2\alpha}_{tt}u,D^{2\alpha}_{xx}u,D^{\alpha}_{t}D^{\alpha}_{x}u...)=0, \ \ \textrm{ $0<\alpha<1$ }
\end{equation}
where $u$ is an unknown functions, $H$ is a polynomial of $u$ and its partial fractional derivatives, and $\alpha$ is order of the mRLd of the function $u=u (x, t)$.

\noindent
The traveling  wave transformation is
\begin{equation} \label{eq1}
\begin{aligned}
& u(x,t)=U(\varepsilon), \\
& \varepsilon=\frac{kx^{\alpha}}{\Gamma(1+\alpha)}-\frac{ct^{\alpha}}{\Gamma(1+\alpha)},
\end{aligned}
\end{equation}
with $k\neq0$ and $c\neq0$ are constants. We use the chain rule
\begin{equation} \label{eq3}
\begin{aligned}
& D^{\alpha}_{t}u=\sigma_{t}\frac{\partial U}{\partial \varepsilon} D^{\alpha}_{t}\varepsilon , \\
& D^{\alpha}_{x}u=\sigma_{x}\frac{\partial U}{\partial \varepsilon} D^{\alpha}_{x}\varepsilon,
\end{aligned}
\end{equation}
with $\sigma_{t},\sigma_{x}$  are sigma indexes \cite{He} and they can be $\sigma_{t}=\sigma_{x}=L$, where $L$ is a constant.

\noindent
Substituting \eqref{eq1} and applying \eqref{eq2} and \eqref{eq3} to \eqref{eq4}, we get following nonlinear ODE
\begin{equation}\label{eq83}
N(U,\frac{dU}{d\varepsilon}, \frac{d^{2}U}{d\varepsilon^{2}}, \frac{d^{3}U}{d\varepsilon^{3}}, ...)=0.
\end{equation}
\section{Modified Kudryashov method}
\noindent
Let the exact solution of \eqref{eq83} can be showed as follows
\begin{equation}\label{eq150}
U(\varepsilon)=a_{0}+a_{1}Q(\varepsilon)+...+a_{N}Q(\varepsilon)^{N},
\end{equation}
where $a_{i}$ values ($i=0,1,2,...,N)$ are arbitrary constants to be found later, but $a_{N}\neq 0$. $Q(\varepsilon)$ has the form 
\begin{equation}\label{eq151}
Q(\varepsilon)=\frac{1}{1+dA^{\varepsilon}}
\end{equation}
which is a solution to the Riccati equation
\begin{equation}\label{eq152}
Q'(\varepsilon )=(Q^2(\varepsilon )-Q(\varepsilon ))lnA
\end{equation}
where $d$ and $A$ are nonzero constants with $A > 0$ and $A\neq 1$. N is revealed by balancing the highest order derivative and nonlinear terms in \eqref{eq83}. Substituting \eqref{eq150} into \eqref{eq83} and comparing the results of the terms with a series of nonlinear equations, new exact solutions will be taken for \eqref{eq4}. 

\section{Applications}
\subsection{Application of ansatz method to space time fractional Klein-Gordon equation}
\noindent
We consider the space-time fractional Klein-Gordon equation of the form 
\begin{equation} \label{eq5}
D^{2\alpha}_{tt}u-a^{2}D^{2\alpha}_{xx}+b^{2}u-\lambda u^3=0, \ \ (t>0, \ \ 0<\alpha\leq 1),\ \
\end{equation}
where $a,b,\lambda$ are constants \cite{Culha}. The bright, dark and singular soliton solutions will be applied to the solitary wave ansatz method.
In order to solve Eq.\eqref{eq5}, using the traveling wave transformation \eqref{eq1}, we obtain to an ODE
\begin{equation} \label{eq6}
L^{2}(a^{2}k^{2}-c^{2})U''-b^{2}U+\lambda U^{3}=0,
\end{equation}
with $U'=\frac{dU}{d\varepsilon}$.
\qquad
\subsubsection{The bright soliton solution}
\noindent
For the bright soliton solution, we let $A,k$ and, $c$ be abritrary constants. Then suppose 
\begin{equation} \label{eq8}
U(\varepsilon)=A\textmd{sech}^{p}(\varepsilon),
\end{equation}
where
\begin{equation} \label{eq9}
\varepsilon=\frac{kx^{\alpha}}{\Gamma(1+\alpha)}-\frac{ct^{\alpha}}{\Gamma(1+\alpha)}.
\end{equation}
It follows from  ansatz \eqref{eq8} and \eqref{eq9} that 
\begin{equation} \label{eq10}
\frac{d^{2}U}{d\varepsilon^{2}}=Ap^{2}\textmd{sech}^{p}(\varepsilon)-Ap(p+1)\textmd{sech}^{p+2}(\varepsilon),
\end{equation}
and
\begin{equation} \label{eq11}
U^{3}=A^{3}\textmd{sech}^{3p}(\varepsilon) .
\end{equation}
Substituting the ansatz \eqref{eq8}-\eqref{eq11} into \eqref{eq6}, the following equation is obtained
\begin{equation} \label{eq12}
\begin{aligned}
& L^{2}(a^{2}k^{2}-c^{2})Ap^{2}\textmd{sech}^{p}(\varepsilon)-L^{2}(a^{2}k^{2}-c^{2})Ap(p+1)\textmd{sech}^{p+2}(\varepsilon)-b^{2}A\textmd{sech}^{p}(\varepsilon) \\
& +\lambda A^{3}\textmd{sech}^{3p}(\varepsilon)=0.
\end{aligned}
\end{equation}
From \eqref{eq12}, we suppose the exponents $p+2$ and $3p$  are equal and from that $p$ is determined as 1. When this value is placed in \eqref{eq12}, it is reduced to the following equation
\begin{equation} \label{eq80}
L^{2}(a^{2}k^{2}-c^{2})A\textmd{sech}(\varepsilon)-2L^{2}(a^{2}k^{2}-c^{2})A\textmd{sech}^{3}(\varepsilon)-b^{2}A\textmd{sech}(\varepsilon) +\lambda A^{3}\textmd{sech}^{3}(\varepsilon)=0.
\end{equation}
From \eqref{eq80}, we obtain the following system of algebraic equations

\begin{displaymath} 
 \left\{ \begin{array}{ll}
\lambda A^{2}-2L^{2}(a^{2}k^{2}-c^{2})=0, \\
L^{2}(a^{2}k^{2}-c^{2})-b^{2}=0.
\end{array} \right.
\end{displaymath} 
\noindent
Solving this system, we get
\begin{equation} \label{eq14}
\begin{aligned}
& A= \mp \sqrt{\frac{2L^{2}(a^{2}k^{2}-c^{2})}{\lambda}}, \\
& c=\mp\sqrt{\frac{L^{2}a^{2}k^{2}-b^{2}}{L^{2}}}.
\end{aligned}
\end{equation}
Finally, we obtain the bright soliton solution for the Fractional Klein-Gordon as follows
\begin{equation} \label{bright1}
u(x,t)= \mp \sqrt{\frac{2L^{2}(a^{2}k^{2}-c^{2})}{\lambda}}  \textmd{sech}\Big(\frac{kx^{\alpha}}{\Gamma(1+\alpha)}\mp\sqrt{\frac{L^{2}a^{2}k^{2}-b^{2}}{L^{2}}}\frac{t^{\alpha}}{\Gamma(1+\alpha)}\Big).
\end{equation}
\begin{figure}
\centering
  \includegraphics[width=16cm]{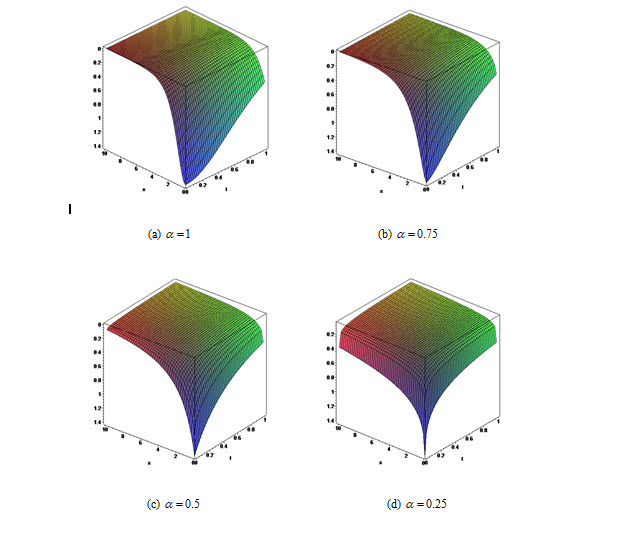}\\
  \caption{The solution $u(x,t)$ for equation \eqref{bright1} when $a=2, k=1, b=1, L=1, \lambda=1$. }
\end{figure}

\noindent
The solution \eqref{bright1} is displayed in Figure 1, in the interval $0<x<10$ and $0<t<1$.

\subsubsection{The dark soliton solution}
\noindent
To obtain dark soliton solution , suppose that
\begin{equation} \label{eq15}
U(\varepsilon)=A\textmd{tanh}^{p}(\varepsilon),
\end{equation}
where
\begin{equation} \label{eq16}
\varepsilon=\frac{kx^{\alpha}}{\Gamma(1+\alpha)}-\frac{ct^{\alpha}}{\Gamma(1+\alpha)},
\end{equation}
which $k,c$ and $A$ are nonzero constant coefficients. From ansatz \eqref{eq15} and \eqref{eq16}, we get
\begin{equation} \label{eq17}
\frac{d^{2}U}{d\varepsilon^{2}}=Ap(p-1)\textmd{tanh}^{p-2}(\varepsilon)-2Ap^{2}\textmd{tanh}^{p}(\varepsilon)+Ap(p+1)\textmd{tanh}^{p+2}(\varepsilon),
\end{equation}
and
\begin{equation} \label{eq18}
U^{3}=A^{3}\textmd{tanh}^{3p}(\varepsilon) .
\end{equation}
Thus, substituting the ansatz \eqref{eq15}-\eqref{eq18} into \eqref{eq6}, it is achieved
\begin{equation} \label{eq19}
\begin{aligned}
& L^{2}(c^{2}-a^{2}k^{2})[Ap(p-1)\textmd{tanh}^{p-2}(\varepsilon)-2Ap^{2}\textmd{tanh}^{p}(\varepsilon)+Ap(p+1)\textmd{tanh}^{p+2}(\varepsilon)] \\
& +b^{2}A\textmd{tanh}^{p}(\varepsilon)-\lambda A^{3}\textmd{tanh}^{3p}(\varepsilon)=0.
\end{aligned}
\end{equation}
From \eqref{eq19}, equating exponents $p$+2 and 3$p$, that gives rise to $p$=1. By using this value, Eq. \eqref{eq19} reduces to
\begin{equation} \label{eq81}
L^{2}(c^{2}-a^{2}k^{2})[-2A\textmd{tanh}(\varepsilon)+2A\textmd{tanh}^{3}(\varepsilon)] +b^{2}A\textmd{tanh}(\varepsilon)-\lambda A^{3}\textmd{tanh}^{3}(\varepsilon)=0.
\end{equation}
From \eqref{eq81}, we find the algebraic system
\begin{equation} \label{eq20}
\begin{aligned}
& 2L^{2}(c^{2}-a^{2}k^{2})-\lambda A^{2}=0, \\
&-2L^{2}(c^{2}-a^{2}k^{2})+b^{2}=0.
\end{aligned}
\end{equation}
Solving the system \eqref{eq20}
\begin{equation} \label{eq21}
\begin{aligned}
& A= \mp \sqrt{\frac{2L^{2}({c^{2}-a^{2}k^{2})}}{\lambda}}, \\
&c=\mp \sqrt{\frac{{b^{2}+2L^{2}a^{2}k^{2}}}{2L^{2}}}.
\end{aligned}
\end{equation}
Finally, we get the dark soliton solution for the Fractional Klein-Gordon as follows:
\begin{equation}\label{dark1}
u(x,t)= \mp \sqrt{\frac{2L^{2}({c^{2}-a^{2}k^{2})}}{\lambda}} \ \  \textmd{tanh}\Big(\frac{kx^{\alpha}}{\Gamma(1+\alpha)}  \mp \sqrt{\frac{{b^{2}+2L^{2}a^{2}k^{2}}}{2L^{2}}}\frac{t^{\alpha}}{\Gamma(1+\alpha)}\Big).
\end{equation}
\begin{figure}
\centering
  \includegraphics[width=16cm]{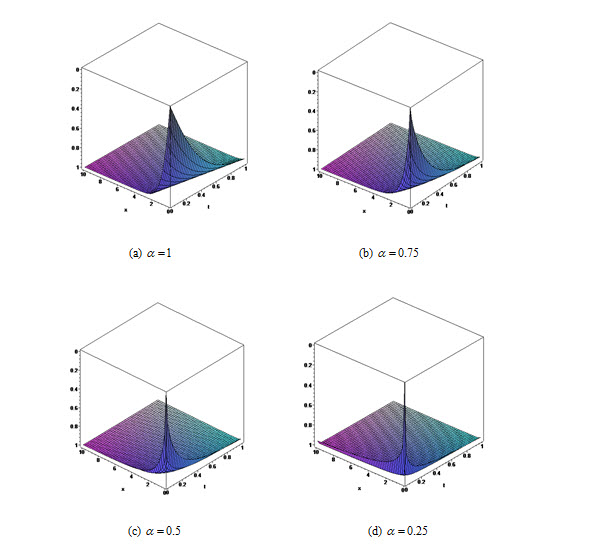}\\
  \caption{The solution $u(x,t)$ for equation \eqref{bright1} when $a=2, k=1, b=1, L=1, \lambda=1$. }
\end{figure}

\noindent
The solution \eqref{dark1} is displayed in Figure 2, in the interval $0<x<10$ and $0<t<1$.

\subsubsection{The singular soliton solution}
\noindent
In finding singular soliton solution we assume 
\begin{equation} \label{eq22}
U(\varepsilon)=A\textmd{csch}^{p}(\varepsilon),
\end{equation}
with
\begin{equation} \label{eq23}
\varepsilon=\frac{kx^{\alpha}}{\Gamma(1+\alpha)}-\frac{ct^{\alpha}}{\Gamma(1+\alpha)},
\end{equation}
where $k,c$ and $A$ are nonzero constant coefficients. From ansatz \eqref{eq22} and \eqref{eq23}, we find
\begin{equation} \label{eq24}
\frac{d^{2}U}{d\varepsilon^{2}}=Ap^{2}\textmd{csch}^{p}(\varepsilon)+Ap(p+1)\textmd{csch}^{p+2}(\varepsilon),
\end{equation}
and
\begin{equation} \label{eq25}
U^{3}=A^{3}\textmd{csch}^{3p}(\varepsilon) .
\end{equation}
Substituting ansatz \eqref{eq22}-\eqref{eq25} into \eqref{eq6}, yields 
\begin{equation} \label{eq26}
\begin{aligned}
& L^{2}(c^{2}-a^{2}k^{2})Ap^{2}\textmd{csch}^{p}(\varepsilon)+L^{2}(c^{2}-a^{2}k^{2})Ap(p+1)\textmd{csch}^{p+2}(\varepsilon)+b^{2}A\textmd{csch}^{p}(\varepsilon) \\
& -\lambda A^{3}\textmd{csch}^{3p}(\varepsilon)=0.
\end{aligned}
\end{equation}
In \eqref{eq26}, when equating exponents $p$+2 and 3$p$, leads $p$=1. Similarly using $p=1$, equation \eqref{eq26} reduces to
\begin{equation} \label{eq82}
L^{2}(c^{2}-a^{2}k^{2})A\textmd{csch}(\varepsilon)+2L^{2}(c^{2}-a^{2}k^{2})A\textmd{csch}^{3}(\varepsilon)+b^{2}A\textmd{csch}(\varepsilon)-\lambda A^{3}\textmd{csch}^{3}(\varepsilon)=0.
\end{equation}
From \eqref{eq82}, we find the algebraic equation system

\begin{displaymath} 
 \left\{ \begin{array}{ll}
2L^{2}(c^{2}-a^{2}k^{2})-\lambda A^{2}=0, \\
L^{2}(c^{2}-a^{2}k^{2})+b^{2}=0.
\end{array} \right.
\end{displaymath} 

\noindent
Solving this system, we get
\begin{equation} \label{eq28}
\begin{aligned}
& A= \mp \sqrt{\frac{2L^{2}(c^{2}-a^{2}k^{2})}{\lambda}} \ (c^{2}-a^{2}k^{2}>0, \lambda < 0),  \\
& c=\mp\sqrt{\frac{L^{2}a^{2}k^{2}-b^{2}}{L^{2}}} \ (L^{2}a^{2}k^{2}-b^{2}>0).
\end{aligned}
\end{equation}
Finally, we find the singular soliton solution for the Fractional Klein-Gordon as follows
\begin{equation} \label{singular1}
u(x,t)= \mp \sqrt{\frac{2L^{2}(c^{2}-a^{2}k^{2})}{\lambda}} \textmd{csch}\Big(\frac{kx^{\alpha}}{\Gamma(1+\alpha)}\mp\sqrt{\frac{L^{2}a^{2}k^{2}-b^{2}}{L^{2}}}\frac{t^{\alpha}}{\Gamma(1+\alpha)}\Big).
\end{equation}
\begin{figure}
\centering
  \includegraphics[width=16cm]{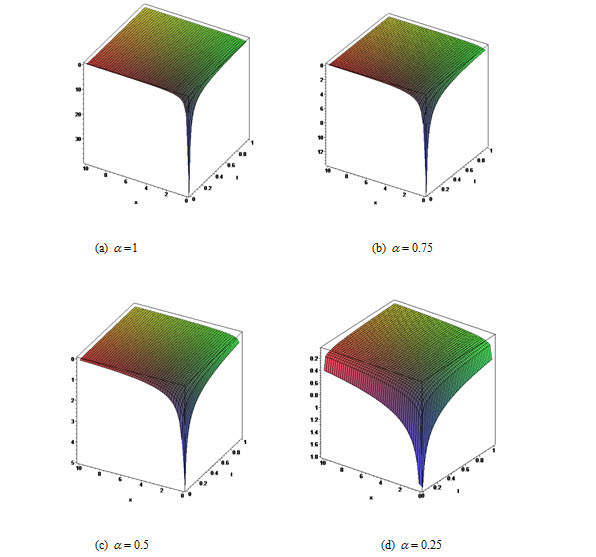}\\
  \caption{The solution $u(x,t)$ for equation \eqref{singular1} when $a=2, k=1, b=1, L=1, \lambda=-1$.}
\end{figure}

\noindent
The solution \eqref{singular1} is displayed in Figure 3, in the interval $0<x<10$ and $0<t<1$.
\qquad
\subsection{Application of modified Kudryashov method to space time fractional Klein-Gordon equation}
\noindent
We consider the space-time fractional Klein-Gordon equation of the form \eqref{eq5}. In order to solve Eq.\eqref{eq5}, using the traveling wave transformation \eqref{eq1}, we obtain to an ODE
\begin{equation} \label{160}
L^{2}(c^{2}-a^{2}k^{2})U''+b^{2}U-\lambda U^{3}=0,
\end{equation}
with $U'=\frac{dU}{d\varepsilon}$. The balance of $U^3$ and $U''$ gives $N=1$. Therefore, we have 
\begin{equation} \label{161}
U(\varepsilon)=a_{0}+a_{1}Q(\varepsilon),\ \ a_{1}\neq0.
\end{equation}
Substituting the solution \eqref{161} and its derivative into \eqref{160} gets 
\begin{equation} \label{eq162}
\begin{aligned}
& \Big(2a_{1}L^{2}(c^{2}-a^{2}k^{2})(lnA)^2-\lambda a_{1}^3 \Big)Q^3(\varepsilon)-3\Big(a_{1}L^{2}(c^{2}-a^{2}k^{2})(lnA)^2+\lambda a_{0}a_{1}^2 \Big)Q^2(\varepsilon) \\
& +\Big(a_{1}L^{2}(c^{2}-a^{2}k^{2})(lnA)^2+b^2a_{1}-3\lambda a_{0}^2a_{1} \Big)Q(\varepsilon)+b^2a_{0}-\lambda a_{0}^3=0.
\end{aligned}
\end{equation}
Equating the coefficients of each power of $Q(\varepsilon)$ and the constant term to zero, solving the resulting system of algebraic equations, we get the following solutions.

\noindent
Case 1:
\begin{equation} \label{163}
a_{0}=-\frac{b}{\lambda \sqrt{\frac{1}{\lambda}}}, \ \ a_{1}=2b\sqrt{\frac{1}{\lambda}}, \ \ c=\mp \sqrt{(lnA)^2a^2k^2L^2+2b^2}
\end{equation}
Substuting \eqref{163} into \eqref{161}, we have 
\begin{equation} \label{164}
U(\varepsilon)=-\frac{b}{\lambda \sqrt{\frac{1}{\lambda}}}+2b\sqrt{\frac{1}{\lambda}}\Big(\frac{1}{1+dA^{\varepsilon}}\Big), \ \ (\lambda>0).
\end{equation}
Finally, we obtain the exact solution of \eqref{eq5}
\begin{equation} \label{165}
u_{1}(x,t)=-\frac{b}{\lambda \sqrt{\frac{1}{\lambda}}}+2b\sqrt{\frac{1}{\lambda}}\Big(\frac{1}{1+dA^{\frac{kx^{\alpha}}{\Gamma(1+\alpha)}\mp \frac{ \sqrt{(lnA)^2a^2k^2L^2+2b^2}t^{\alpha}}{\Gamma(1+\alpha)}}}\Big), \ \ (\lambda>0).
\end{equation}
\noindent
Case 2:
\begin{equation} \label{166}
a_{0}=\frac{b}{\lambda \sqrt{\frac{1}{\lambda}}}, \ \ a_{1}=-2b\sqrt{\frac{1}{\lambda}}, \ \ c=\mp \sqrt{(lnA)^2a^2k^2L^2+2b^2}
\end{equation}
Substuting \eqref{166} into \eqref{161}, we get 
\begin{equation} \label{167}
U(\varepsilon)=\frac{b}{\lambda \sqrt{\frac{1}{\lambda}}}-2b\sqrt{\frac{1}{\lambda}}\Big(\frac{1}{1+dA^{\varepsilon}}\Big), \ \ (\lambda>0).
\end{equation}
Finally, we obtain the exact solution of \eqref{eq5}
\begin{equation} \label{168}
u_{2}(x,t)=\frac{b}{\lambda \sqrt{\frac{1}{\lambda}}}-2b\sqrt{\frac{1}{\lambda}}\Big(\frac{1}{1+dA^{\frac{kx^{\alpha}}{\Gamma(1+\alpha)}\mp \frac{\sqrt{(lnA)^2a^2k^2L^2+2b^2}t^{\alpha}}{\Gamma(1+\alpha)}}}\Big), \ \ (\lambda>0).
\end{equation}

\section{Conclusion}
In this article, the space time-fractional Klein-Gordon equation with cubic nonlinearities is investigated for soliton and exact solutions. Complex fractional transformation is utilized to  attain the nonlinear ODE from  this equation. Bright, dark and singular soliton solutions are obtained with solitary wave ansatz method. Moreover, some exact solutions are found with modified Kudryashov method. The results are proof that these methods are accurate and effective. In addition, graphs of all soliton solutions are drawn for the appropriate coefficients.

%
%
%
%
%
%


\begin{thebibliography}{9}

\bibitem{Miller} Miller, K. S., Ross, B., {\it An Introduction to the Fractional Calculus and Fractional Differential Equations}, Wiley, New York (1993)

\bibitem{Podlubny} Podlubny, I., {\it Fractional Differential Equations} , Academic Press, California (1999)

\bibitem{Kilbas}  Kilbas, A. A., Srivastava, H. M., Trujillo, J. J., {\it Theory and Applications of Fractional Differential Equations}, Elsevier, Amsterdam (2006)

\bibitem{Ray} Ray, Saha S., {\it New exact solutions of nonlinear fractional acoustic wave equations in ultrasound}, Comput Math Appl., 71 (2016) 859-868

\bibitem{Taghizadeh} Taghizadeh, N., Najand, F., M., Soltani Mohammadi V., {\it New exact solutions of the perturbed nonlinear fractional Schrödinger equation using two reliable methods}, Appl Math., 10 (2015) 139-148.

\bibitem{Mirzazadeh} Mirzazadeh, M., Eslami, M., Biswas, A., {\it Solitons and periodic solutions to a couple of fractional nonlinear evolution equations}, Pramana J Phys., 82 (2014) 465-476.

\bibitem{Eslami} Eslami, M., Rezazadeh, H.,{\it The first integral method for Wu-Zhang system with conformable time-fractional derivative}, Calcolo., 53 (2016) 475-485.

\bibitem{Cenesiz} Cenesiz, Y., Baleanu, D., Kurt, A., et al. {\it New exact solutions of Burgers' type equations with conformable derivative}, Waves Random Complex Media., 27 (2017) 103-116.

\bibitem{Guner} G\"{u}ner, \"{O}., Bekir, {\it A. Exact solutions of some fractional differential equations arising in mathematical biology}, Int J Biomath., 8 (2015) 1550003-1 - 1550003-17.

\bibitem{Bekir1} Bekir, A., G\"{u}ner, \"{O}., \c{C}evikel A., C., {\it Fractional complex transform and exp-function methods for fractional differential equations}, Abstr Appl Anal. (2013) ,Article ID 426462.

\bibitem{Bekir2} Bekir, A., G\"{u}ner, \"{O}., Bhrawy, A., H., et al. {\it Solving nonlinear fractional differential equations using expfunction and (G'/G) expansion methods}, Rom J Phys., 60 (2015) 360-378.

\bibitem{Bin} Bin Z., {\it (G'/ G)-expansion method for solving fractional partial differential equations in the theory of mathematical physics}, Commun Theor Phys., 58 (2012) 623-630.

\bibitem{Bekir3} Bekir, A., G\"{u}ner, \"{O}., {\it Exact solutions of nonlinear fractional differential equations by (G'/ G)-expansion method}, Chin Phys B., 22 (2013) 110202-1 - 110202-6.

\bibitem{Baleanu} Baleanu, D., U\v{g}urlu, Y., {\it Inc M, et al. Improved (G'/ G)-expansion method for the time-fractional biological population model and Cahn.Hilliard equation}, J. Comput Nonlinear Dyn., 10 (2015) 051016.1 - 051016-8.

\bibitem{Aksoy} Aksoy, E., \c{C}evikel A.,C., Bekir, A., {\it Soliton solutions of (2+1)-dimensional time-fractional Zoomeron equation}, Optik., 127 (2016) 6933-6942.

\bibitem{Bekir4} Bekir, A., Aksoy, E., {\it Application of the subequation method to some differential equations of time fractional order}, Rom J Phys., 10 (2015) 054503-1 - 054503-5.

\bibitem{Matinfar} Matinfar, M., Eslami, M., Kordy, M., {\it The functional variable method for solving the fractional Korteweg de Vries equations and the coupled Korteweg de Vries equations}, Pramana J Phys. 85 (2015) 583-592.

\bibitem{Guner1} G\"{u}ner, \"{O}., Eser, D., {\it Exact solutions of the space time fractional symmetric regularized long wave equation using different methods}, Adv Math Phys., (2014) Article ID 456804.

\bibitem{Bulut} Bulut, H., Baskonus, H.,M., Pandir, Y., {\it The modified trial equation method for fractional wave equation and time fractional generalized Burgers equation}, Abstr Appl Anal.,(2013) Article ID 636802.

\bibitem{Odabasi} Odabas{\i}, M., M{\i}s{\i}rl{\i}, E., {\it On the solutions of the nonlinear fractional differential equations via the modified trial equation method}, Math Methods Appl Sci. (2015) DOI:10.1002/mma.3533

\bibitem{Biswas1} Biswas, A., {\it 1-Soliton solution of the K(m,n) equation with generalized evolution}, Phys. Lett. A,
372 (2008), 4601-4602.

\bibitem{Karishnan} Krishnan, E.,V., Biswas, A,. {\it Solutions to the Zakharov Kuznetsov equation with higher order nonlinearity by mapping and ansatz methods}, Phys. Wave Phenom. 18 (2010) 256-261.

\bibitem{Bekir} Bekir, A., Aksoy, E. and Guner, O., {\it Optical soliton solutions of the Long-Short-Wave interaction
system}, Journal of Nonlinear Optical Physics and Materials, 22(2) (2013), 1350015(11 pages).

\bibitem{Bekir5}  Bekir, A. and Guner, O., {\it Bright and dark soliton solutions of the (3 + 1)-dimensional generalized
Kadomtsev-Petviashvili equation and generalized Benjamin equation}, Pramana-J. Phys., 81(2) (2013), 203-214.

\bibitem{Bekir6} Bekir, A. and Guner, O., {\it Topological (dark) soliton solutions for the Camassa-Holm type equations},
Ocean Eng., 74 (2013), 276-279.

\bibitem{Biswas} Biswas, A., {\it 1-Soliton solution of the B(m,n) equation with generalized evolution}, Commun.Nonlinear Sci.
Numer. Simul., 14 (2009), 3226-3229.

\bibitem{Biswas2} Biswas, A., {\it Optical solitons with time-dependent dispersion, nonlinearity and attenuation in a
power-law media}, Commun. Nonlinear Sci. Numer. Simulat., 14 (2009), 1078-1081.

\bibitem{Biswas3} Biswas, A., and Milovic,D., {\it Bright and dark solitons of the generalized nonlinear Schr\"{o}dingers
equation}, Commun. Nonlinear Sci. Numer. Simulat., 15 (2010), 1473-1484.

\bibitem{Biswas4} Biswas, A., Triki,H., Hayat, T., and  Aldossary, O. M. {\it 1-Soliton solution of the generalized Burgers
equation with generalized evolution}, Applied Mathematics and Computation, 217 (2011), 10289-
10294.

\bibitem{Triki} Triki,H., and  Wazwaz, A. M., {\it  Bright and dark soliton solutions for a K(m,n) equation with
t-dependent coefficients}, Phys. Lett. A, 373 (2009), 2162-2165.

\bibitem{Khalique} Khalique, C. M.  and  Biswas,A., {\it Optical solitons with parabolic and dual-power law nonlinearity
via Lie group analysis}, Journal of Electromagnetic Waves and Applications, 23(7) (2009), 963-973.

\bibitem{Guner2} G\"{u}ner, \"{O}., {\it Singular and non-topological soliton solutions for nonlinear fractional differential equations}, Chinese Physics B, 24 (2015) 100201

\bibitem{Mirzazadeh1} Mirzazadeh, M., {\it Topological and non-topological soliton solutions to some time-fractional differential equations}, Pramana-J. Phys. 85 (2015) 17-29

\bibitem{Kudryashov} Kudryashov, N. A. {\it One method for finding exact solutions of nonlinear differential equations}, Commun. Nonlinear Sci. Numer. Simulat. 17 (6) (2012) 2248–2253.

\bibitem{Ege} Ege, S.M.,  Misirli, E. {\it The modified Kudryashov method for solving some fractional-order nonlinear equations}, Adv. Difference Equ. 2014 (2014) 135.

\bibitem{Korkmaz} Korkmaz, A. {\it Exact solitons to (3+1) conformable time fractional jimbo Miwa, Zakharov Kuznetsov and modified Zakharov Kuznetsov equations}, Commun.Theor. Phys. 67 (2017) 479–482.

\bibitem{Hosseini} Hosseini, K., Ayati, Z., {\it Exact solutions of space-time fractional EW and modified EW equations using Kudryashov method}, Nonlinear Sci Lett A., 7 (2016) 58-66.

\bibitem{Saha} Saha, R.,S., {\it New analytical exact solutions of time fractional KdV KZK equation by Kudryashov methods}, Chin Phys B., 25 (2016) 040204-1 - 040204-7.

\bibitem{Wazwaz} Wazwaz, AM. {\it Compactons, solitons and periodic solutions for some forms of nonlinear Klein–
Gordon equations}, Chaos, Solitons Fractals., 28 (2006) 1005–1013.

\bibitem{biswasA} Biswas, A. Zony, C., Zerrad, E. {\it Soliton perturbation theory for the quadratic nonlinear Klein–Gordon equation},  Applied Mathematics and Computation, 203 (2008) 153.

\bibitem{Alireza} Golmankhaneh A., K., Golmankhaneh, A., Baleanu, D., {\it On nonlinear fractional Klein-Gordon equation}. Signal Processing, 91 (2011) 446-51.

\bibitem{Tamsir} Tamsir, M., Srivastava, V., {\it Analytical study of time-fractional order Klein-Gordon equation}, Alexandria Engineering Journal 55 (2016) 561-567

\bibitem{Hosseini1} Hosseini, K., Mayeli, P., Ansari, R., {\it Modified Kudryashov method for solving the conformable time-fractional Klein-Gordon
equations with quadratic and cubic nonlinearities}, Optik, 130 (2017) 737-742.

\bibitem{Unsal} Unsal, O., Guner, O., Bekir, A., {\it Analytical approach for space-time fractional Klein-Gordon equation}, Optik 135 (2017) 337-345

\bibitem{Hosseini2} Hosseini, K., Mayeli, P., Ansari, R., {\it Bright and singular soliton solutions of the conformable time-fractional Klein-Gordon equations with different
nonlinearities}, Waves in Random and Complex Media 28 (2018) 426-434

\bibitem{Culha} \c{C}ulha, S., Da\d{s}c{\i}o\~{g}lu, A., {\it Analytic solutions of the space time conformable fractional
Klein Gordon equation in general form}, Waves in Random and Complex Media 29 (2019) 775-790

\bibitem{Shallal} Shallal, M.,A.,Jabbar, H.,N.,  Ali, K., K., {\it Analytic solution for the space-time fractional Klein-Gordon and coupled
conformable Boussinesq equations}, Results in Physics 8 (2018) 372-378

\bibitem{Jumarie} Jumarie, G., {\it Modified Riemann-Liouville derivative and fractional Taylor series of nondifferentiable
functions further results}, Comput. Math. Appl., 51 (2006) 1367-1376

\bibitem{Jumarie1} Jumarie, G., {\it Table of some basic fractional calculus formulae derived from a modified
Riemann-Liouville derivative for nondifferentiable functions}, Appl. Maths. Lett., 22 (2009) 378-385

\bibitem{He} He, J. H.,  Elegan,S. K., and  Li, Z. B., {\it Geometrical explanation of the fractional complex transform
and derivative chain rule for fractional calculus}, Physics Letters A, 376 (2012), 257-259.

\end{thebibliography}
\end{document}